# The phonon dispersion relation of a Bose-Einstein condensate

I. Shammass,[1] S. Rinott,[2] A. Berkovitz,[2] R. Schley,[2] and J. Steinhauer[2]

[1]*Department of Condensed Matter Physics, Weizmann Institute of Science, Rehovot 76100, Israel*

[2]*Department of Physics, Technion—Israel Institute of Technology, Technion City, Haifa 32000, Israel*

We measure the oscillations of a standing wave of phonons in a Bose-Einstein condensate, thus obtaining the dispersion relation. We present the technique of short Bragg pulses, which stimulates the standing wave. The subsequent oscillations are observed *in situ*. It is seen that the phonons undergo a 3D to 1D transition, when their wavelength becomes longer than the *transverse* radius of the condensate. The 1D regime contains an inflection point in the dispersion relation, which should decrease the superfluid critical velocity according to the Landau criterion. The inflection point also represents a minimum in the group velocity, although the minimum is not deep enough to result in a roton. The 3D-1D transition also results in an increase in the lifetime of the standing-wave oscillations, and a breakdown of the local density approximation. In addition, the static structure factor is measured in the long-wavelength regime. The measurements are enabled by the high sensitivity of the new technique.

The excitation spectrum gives the energy of a quasiparticle [1] as a function of the momentum, whereas the dispersion relation gives the oscillation frequency as a function of the wavenumber, emphasizing the wave nature of the phonons. While the excitation spectrum of a Bose-Einstein condensate has been studied extensively [2-6], the dispersion relation has not been measured. Previous measurements have employed Bragg spectroscopy [2,3], in which the condensate absorbs a pre-determined momentum. The energy is adjusted to find the resonance condition, thus obtaining a point on the excitation spectrum [4]. On the other hand, the energy can be determined by tomographic imaging [5]. In either case, the results were well-described by the local density approximation (LDA) [2,7]. In this approximation, each point in the condensate is modeled as a homogeneous condensate, and a suitable average is taken over the inhomogeneous density of the actual condensate. Going beyond the LDA, various radial modes were resolved [6,8]. The present work introduces an alternative technique to Bragg spectroscopy, in which phonons are created with a pre-determined momentum, and are allowed to freely evolve. The oscillation frequency is observed, thus obtaining a point on the dispersion relation. This technique is particularly suitable for long-wavelength phonons. This technique is thus complementary to Bragg spectroscopy, which gives the excitation spectrum rather than the dispersion relation, and is particularly suitable for short wavelengths.

In order to create the phonons, we present the technique of short Bragg pulses. The pulse employed is similar to that used in the Kapitza-Dirac effect [9], a phenomenon involving single particles rather than phonons. Two far-detuned beams with a relative angle $\theta$ impinge on the condensate for a short time $\tau$, as shown in Fig. 1(a). The photons in the beams have a large energy uncertainty $\hbar\delta\omega$ which is on the order of $\hbar/\tau$. This allows the condensate to absorb a photon from either beam and emit a photon into the other, creating a phonon with energy $\hbar\omega_k$, as long as $\omega_k \ll 1/\tau$. The wavenumber $k$ is precisely determined by $\theta$. Counterpropagating phonons with well-defined $k$ are thus produced, resulting in a standing wave.

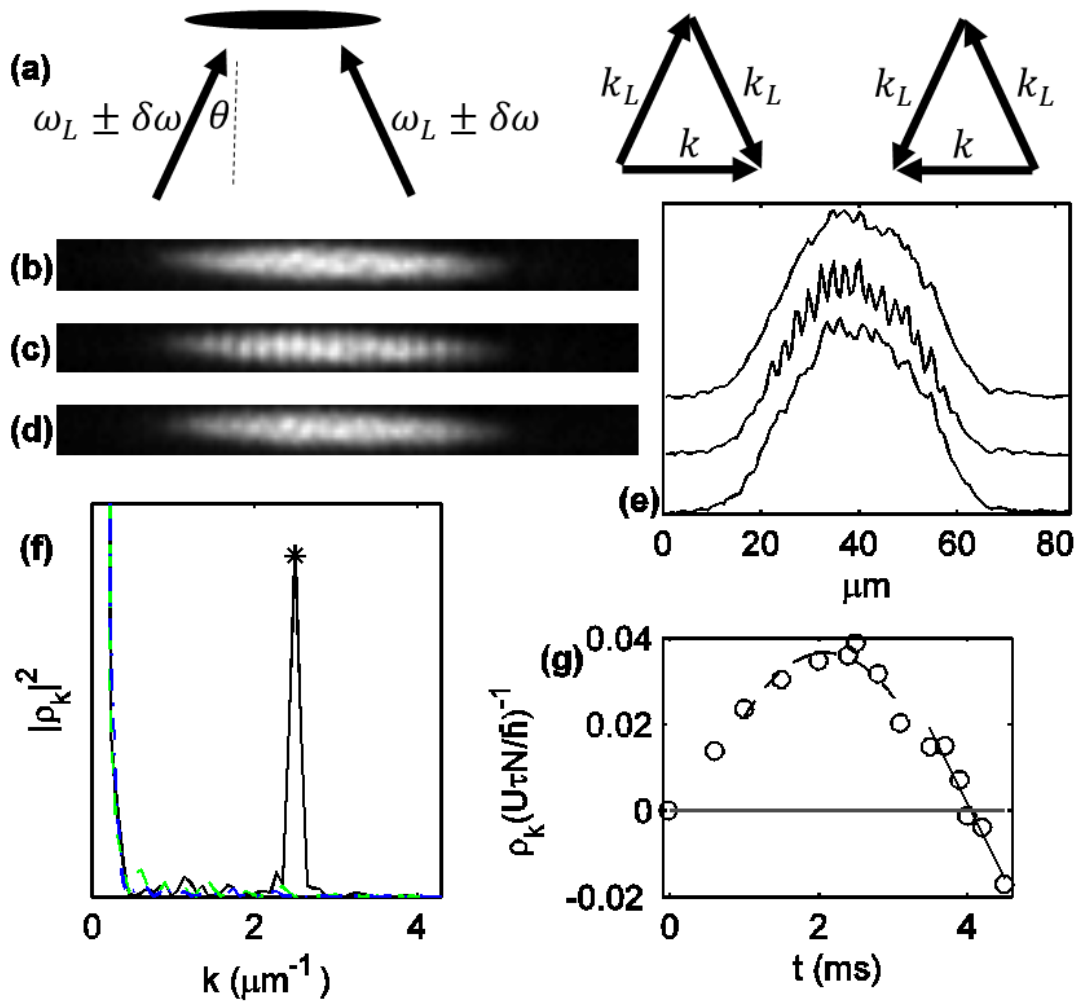

FIG. 1. Creating a phonon standing wave by short Bragg pulses. The larger condensate is shown. (a) Two far-detuned laser beams, with frequency $\omega_L$ and wavenumber $k_L$, impinge on the condensate. Absorption from the right (left) beam and emission into the left (right) beam results in the production of a right-moving (left-moving) phonon with wavenumber $k$. The combination of the left and right-moving phonons results in a standing wave. (b-d) Phase-contrast images of the *in situ* condensate, for a short 22 μsec Bragg pulse with $k = 2.47$ μm$^{-1}$. (b) shows the time just before the pulse. (c) shows the first antinode of the standing wave, at 250 μsec. (d) shows the first node, at 510 μsec. (e) Integrated profiles

of the images. The top, middle, and bottom curves correspond to (b), (c), and (d), respectively. The top and middle curves have been shifted vertically for clarity. (f) The magnitude squared of the Fourier transform of the profiles. The blue dash-dotted, black solid, and green dashed curves correspond to before the pulse, the antinode, and the node, respectively. The asterisk indicates the $k$-value corresponding to the standing wave. (g) The time dependence of the Fourier transform of the standing wave, for $k = 0.35\ \mu m^{-1}$. The linear fit at the zero crossing determines $\omega_k$ (solid line). The parabolic fit at the maximum determines $S_o(k)$ (dashed curve).

During a short Bragg pulse, the energy in the phonons increases as $dE/dt = U^2 t k^2 N/2m$, where $U$ is the amplitude of the sinusoidal potential resulting from the interference between the two laser beams, $N$ is the number of atoms in the condensate, and $m$ is the atomic mass [10]. This expression is independent of the frequency difference between the beams, due to the shortness of the pulse. Writing the energy in terms of quasiparticle number, one obtains $N_k = N_{-k} = (U\tau/\hbar)^2 S_o(k) N/4$, and $S_o(k) = \hbar k^2/2m\omega_k$ is the zero-temperature static structure factor.

After the short Bragg pulse, the phonons freely propagate in the condensate. The wavefunction of a homogeneous condensate in the presence of phonons is given by [10,11]

$$\psi = (\psi_o + \delta\psi)e^{-i\frac{\mu}{\hbar}t} \tag{1}$$

where $\mu$ is the chemical potential, and the small perturbation is given by

$$\delta\psi = \sum_{\mathbf{k}} \sqrt{\frac{N_{\mathbf{k}}}{V}} \left[ u_k e^{i(\mathbf{k}\cdot\mathbf{r}-\omega_k t)} + v_k e^{-i(\mathbf{k}\cdot\mathbf{r}-\omega_k t)} \right]$$

where $V$ is the volume of the condensate, and $u_k$ and $v_k$ are the Bogoliubov amplitudes, which were measured in Ref. 12. We find that the density is given by

$$n = |\psi|^2 = n_o \left[ 1 + \frac{2}{\sqrt{N}} \sum_{\mathbf{k}} \sqrt{N_{\mathbf{k}} S_o(k)} \cos(\mathbf{k}\cdot\mathbf{r} - \omega_k t) \right] \quad (2)$$

where $n_o$ is the average density. For the standing wave, the Fourier transform of the density is given by

$$\rho_k = \left(\frac{U\tau}{\hbar}\right) N S_o(k) \sin(\omega_k t). \quad (3)$$

The standing wave is at a node just after the short Bragg pulse ($t = 0$), and we have adjusted the origin in time accordingly. Thus, by observing the frequency and amplitude of the oscillation of $\rho_k$, we obtain the dispersion relation $\omega_k$, as well as $S_o(k)$.

The sinusoidal potential is created by imaging a spatial light modulator (SLM) onto the condensate. The SLM is illuminated by a far-detuned laser (803.5 nm). The image is filtered in the Fourier plane [13], resulting in two Bragg beams whose angle can be varied by changing the image on the SLM. The condensate is composed of $^{87}$Rb atoms in the $F = 2$, $m_F = 2$ state and is confined in a cylindrically-symmetric harmonic magnetic potential, with radial and axial frequencies of $\omega_\perp = 224$ Hz and $\omega_z = 26$ Hz, respectively. We primarily study a larger condensate with $\mu/h = 2340$ Hz, as well as a smaller condensate with $\mu/h = 1200$ Hz. Immediately after the short Bragg pulse, the condensate appears unperturbed, as seen in Fig.

1(b), and in the integrated profile, Fig. 1(e). After a time corresponding to one-fourth of a period however, the density modulation is at a maximum, as seen in Fig. 1(c). The density modulation disappears again after half a period, as seen in Fig. 1(d). In Fig. 1(b)-(e), the only potential present is the harmonic trapping potential.

Fig. 1(f) shows the magnitude squared of the Fourier transform of the profiles. The asterisk indicates the point corresponding to the applied $k$. We plot one phase component of this point as a function of time, as shown in Fig. 1(g). The phase is chosen to be that of the density modulation near the anti-node (the solid black curve of Fig. 1(f)). Since this measurement technique employs a pre-determined spatial frequency $k$ and is phase-sensitive, it has the advantages of a lock-in amplifier [14]. Furthermore, the small perturbation term $\delta\psi$ in (1) interferes with the large condensate term $\psi_o$, enhancing the signal by a factor of $\sqrt{N}$.

The time of the zero crossing in Fig. 1(g) gives $\pi/\omega_k$ by (3). This time is determined by a linear fit to the data in the region of the zero crossing. Repeating the experiment for many values of $k$ gives the dispersion relation $\omega_k$, as shown in Fig. 2 for both values of $\mu$. The error bars are too small to be seen for most points. Indeed, the errors are an order of magnitude smaller than the result for Bragg spectroscopy [4]. In addition to the advantages discussed above relative to Bragg spectroscopy, here the frequency is found by a zero crossing rather than finding the center of a broad peak in the response function. The black curves of Fig. 2 are the result of a 2D simulation of the Gross-Pitaevskii equation (GPE). This simulation is cylindrically symmetric with radial and axial coordinates. The agreement between the measured values and the simulation is excellent. The lowest measured frequency is seen to be 80 Hz. Since this is much greater than the axial trap frequency of 26 Hz, discrete axial modes can be neglected [15]. Fig. 2 also shows the local density approximation (LDA), which agreed well with previous measurements of the excitation energy [2-5].

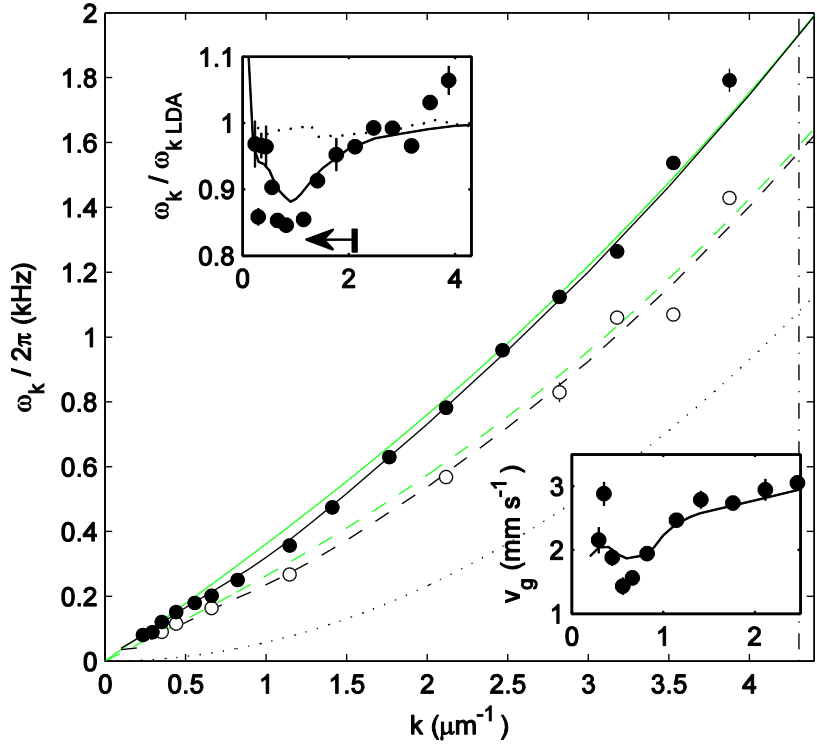

FIG. 2. The phonon dispersion relation of a Bose-Einstein condensate. The filled (open) circles indicate $\mu/h = 2340$ Hz (1200 Hz). The error bars (which are too small to be seen for most points) indicate the standard error of the mean. The black curves indicate the result of the 2D GPE simulation. The green curves are the LDA approximation. Solid (dashed) curves correspond to the larger (smaller) value of $\mu$. The dotted curve is the free-particle spectrum $\omega_k = \hbar k^2/2m$. The dash-dotted line indicates the inverse

healing length, the maximum $k$ of the phonon regime, for the larger value of $\mu$. The upper inset shows the ratio between the dispersion relation and the LDA approximation for the experiment (filled circles), the 2D GPE simulation (solid curve), and the 1D GPE simulation (dotted curve), for the larger value of $\mu$. The arrow indicates the 1D regime. The lower inset shows the group velocity for the experiment (filled circles) and the 2D GPE simulation (solid curve), for the larger value of $\mu$.

In Fig. 2 it is seen that both the measured and simulated dispersion relations are depressed relative to the LDA curve near $k = 1$ μm$^{-1}$. This is particularly visible for the larger value of $\mu$. This is emphasized in the inset, which shows the ratio between the dispersion relations. The ratio is clearly below unity in this $k$-regime. Here, the depression in the dispersion relation is below a 3D-1D transition which occurs when the wavelength becomes longer than the *transverse* radius of the condensate. This result can be understood in terms of the discretization of the modes in the transverse (radial) direction. The broad LDA frequency spectrum is actually divided into narrow radial modes [6,8]. For small $k$, the LDA linewidth becomes sufficiently narrow that only a single radial mode is excited [6], giving the phonons a 1D character. Equating the LDA linewidth $\delta\omega_k \approx 0.6\omega_k$ (FWHM) [7] with the spacing between radial modes of approximately $2\omega_\perp$ [6,8], a single mode will be excited for

$$\lambda > 0.9R_\perp \qquad (4)$$

where $\lambda$ is the phonon wavelength, and $R_\perp$ is the Thomas-Fermi radius of the condensate perpendicular to the direction of propagation. The single mode excited in this frequency regime does not necessarily have the same frequency as the average over the LDA lineshape. Thus, the measured dispersion relation deviates from the LDA curve. Not surprisingly, the depression in the dispersion relation can be seen in simulations of the lowest radial mode [6,8]. The inequality (4) is indicated by the arrow in the upper inset of Fig. 2, which is seen to agree with the regime of the depression in the dispersion relation. As an additional verification that the transition is associated with the transverse degree of freedom, we perform a 1D GPE simulation of the experiment. We compare the results of this simulation with a 1D LDA average, as indicated by the dotted line in the upper inset of Fig. 2. It is seen that the simulation always agrees with the LDA, since there are no transverse modes present.

According to the Landau criterion [1], the depression in the dispersion relation corresponds to a slight decrease in the superfluid critical velocity. At the depression, $\omega/k = 1.91 \pm 0.01$ mm sec$^{-1}$, which is the lowest value for the entire measured dispersion relation, and thus corresponds to the critical velocity. For the LDA, $\omega/k$ is smallest for $k = 0$, and takes the value 2.22 mm sec$^{-1}$. Thus, the critical velocity is suppressed to 0.9 of its LDA value, due to the 1D nature of the phonons. The depression also implies a minimum in the group velocity $v_g = d\omega/dk$, as seen in the lower inset of Fig. 2. In superfluid helium, the minimum in the group velocity corresponds to a roton [1,16], an excitation whose wavelength is on the order of the effective hard-sphere radius of the atoms [17]. Here the minimum is weaker, and the relevant length scale is the transverse radius of the condensate.

The oscillation amplitude in Fig. 1g gives $S_o(k)$ by (3). This value is determined by a parabolic fit in the region of the maximum. The result is shown in Fig. 3. The experimental values in Fig. 3 have been multiplied by a factor of 1.5, in order to agree with the 2D GPE simulation for low $k$. This factor is probably required due to uncertainty in the intensity of the Bragg beams at the location of the condensate, which affects $U$ in (3). For small $k$, the functional agreement with the simulation is very good. The uncertainty expressed by the error bars is seen to be an order of magnitude improvement relative to the Bragg spectroscopy results in this regime [4]. However, for larger $k$, the measured values roll off due to the finite resolution of the imaging system, which decreases the apparent values of $\rho_k$.

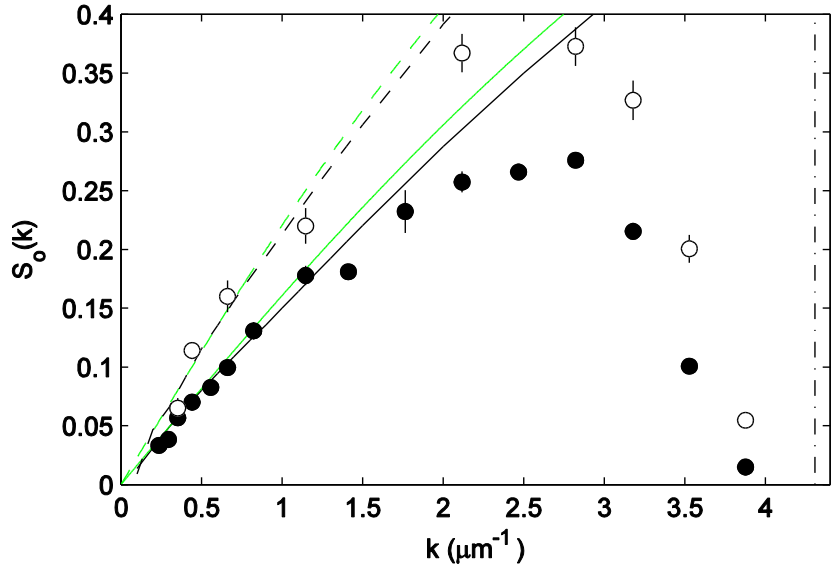

FIG. 3. The static structure factor. The filled (open) circles indicate $\mu/h = 2340$ Hz (1200 Hz). The error bars indicate the standard error of the mean. The black curves indicate the result of the 2D GPE simulation. The green curves are the LDA approximation. Solid (dashed) curves correspond to the larger (smaller) value of $\mu$. The dash-dotted line indicates the inverse healing length for the larger value of $\mu$.

We now explore the decay of the oscillations, as seen in Fig. 4(a). The short Bragg pulse (as with any Bragg pulse) excites several Bogoliubov modes [6,8] which subsequently dephase, causing the decay. The filled circles of Fig. 4(b) show the $Q$ factor of the oscillations; the number of radians $\omega_k t$ required for the amplitude to decay by $1/\sqrt{e}$. It is seen that for long wavelengths, $Q$ increases significantly. This is in contradiction with the LDA prediction for the decay, which is due to the dephasing of the oscillations at various locations in the condensate. For small $k$, the LDA linewidth implies a constant $Q$ of $1/0.6$, as indicated by the dashed line in Fig. 4(b). We can make a more precise LDA prediction by extending the LDA to include the time-dependence of the density. Writing (2) for the standing wave, and taking the various quantities to be functions of position,

$$n(\mathbf{r},t)_{\text{LDA}} = n_o(\mathbf{r})\left[1 + \left(\frac{Vt_B}{\hbar}\right)S_o(\mathbf{r},k)\{\sin[kx - \omega_k(\mathbf{r})t] - \sin[kx + \omega_k(\mathbf{r})t]\}\right] \qquad (4)$$

where $x$ is the direction of oscillation of the standing wave, and we have adjusted the time origin to form a node at $t = 0$. By (4), each point in the condensate oscillates with a different frequency and amplitude, depending on the local density. The density profile and its Fourier transform $\rho_k$ is computed for (4), yielding the solid curve in Fig. 4(b). Again it is seen that the measured $Q$ significantly exceeds the LDA prediction for small $k$. The discrepancy between the measurement and the LDA is due to the 3D-1D transition. In the 1D regime, the single radial mode has a much narrower linewidth than the LDA, resulting in larger $Q$. This regime is indicated by the arrow in Fig. 4(b). It is seen that the arrow approximately corresponds to the region in which the measured values display increased $Q$.

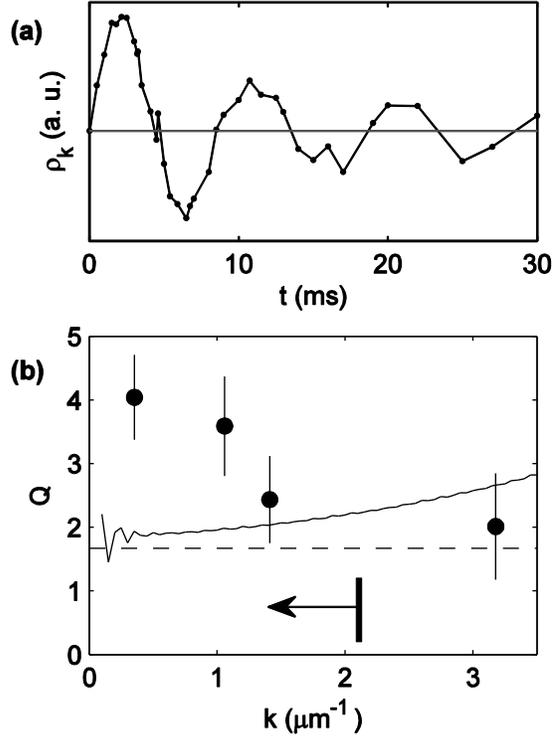

FIG. 4. The decay of the phonon standing wave. The larger condensate is shown. (a) The amplitude of the standing wave as a function of time, for $k = 0.35$ μm$^{-1}$. (b) The $Q$ of the oscillation. The error bars reflect the uncertainty due to the finite number of cycles measured. The solid curve is the LDA result, by (4). The dashed line is the LDA result for small $k$. The arrow indicates the 1D regime.

In conclusion, we have studied the time evolution of phonons in a Bose-Einstein condensate, thus obtaining the dispersion relation and the static structure factor. The phonon standing wave is stimulated by the novel technique of short Bragg pulses. The sensitivity of the measurement is an order of magnitude beyond the previous state-of-the-art, which allows for the observation of a 3D-1D transition. In the 1D regime corresponding to long wavelengths, the phonons are characterized by a single radial mode. This mode has a lower frequency and longer lifetime than predicted by the local density approximation. The 1D regime results in a decreased superfluid critical velocity, and a minimum in the group velocity. The 1D nature of the long-wavelength modes is important for the analysis of sonic black holes [18-23].

We thank Chris Westbrook for helpful comments. This work was supported by the Israel Science Foundation.